\documentclass[st,twocolumn]{jpsj3}
\usepackage{txfonts}
\usepackage{graphicx}
\usepackage{dcolumn}
\usepackage{bm}
\usepackage{color}
\usepackage{comment}
\usepackage{overcite}

\title{Topological Semimetals Studied by Ab Initio Calculations }
\author{Motoaki Hirayama$^{1,2,3}$\thanks{motoaki.hirayama@riken.jp}, 
Ryo Okugawa$^{1}$, and Shuichi Murakami$^{1,2,4}$\thanks{murakami@stat.phys.titech.ac.jp}}
\inst{$^1$Department of Physics, Tokyo Institute of Technology, 2-12-1 Ookayama, 
Meguro-ku, Tokyo 152-8551, Japan \\
$^2$TIES, Tokyo Institute of Technology, 2-12-1 Ookayama, 
Meguro-ku, Tokyo 152-8551, Japan \\
$^3$RIKEN Center for Emergent Matter Science,
Hirosawa 2-1, Wako, Saitama 351-0198, Japan \\
$^4$CREST, Tokyo Institute of Technology, 2-12-1 Ookayama, 
Meguro-ku, Tokyo 152-8551, Japan \\
} 
\abst{
In topological semimetals such as Weyl, Dirac, and nodal-line semimetals, 
the band gap closes at points or along lines in ${\bf k}$ space
which are not necessarily located at high-symmetry positions in the Brillouin zone.
Therefore, it is not straightforward to find these topological semimetals by
 {\it ab initio} calculations because the band structure is usually calculated 
only along high-symmetry lines. 
In this paper, we review recent studies on 
topological semimetals by {\it ab initio} calculations. We explain theoretical frameworks
which can be used for the search for topological semimetal materials, and some 
numerical methods used in the {\it ab initio} calculations. 
}

\begin{document}
\maketitle

\section{Introduction}


Theoretical proposals of  a topological insulator  (TI) phase 
\cite{Kane05a,Kane05b,BZ,Hasan,Qi1,BYan-review} have triggered intensive studies on various types of topological states. In addition to a TI phase, there are various topological phases with a gap, such as an integer  quantum Hall system \cite{Klitzing}, and a topological crystalline insulator \cite{Fu11}.
They have characteristic surface or edge states, which manifest the
topological nature of the bulk wave functions.
Through the research on TIs, a new category of topological phases,
called  topological semimetals  (topological metals), has been found in condensed matter physics. 
Degeneracy between
 conduction and valence bands usually 
occurs only at high-symmetry points or lines in ${\bf k}$ space. 
In contrast, research on 
topological semimetals has shown us other possibilities for band degeneracies that originate from topology. 
In  such topological semimetals, the 
band gap closes at generic ${\bf k}$ points, and this closing of the gap  
originates from topological reasons. 

\begin{figure}[h]
\begin{center}
\includegraphics[width=85mm]{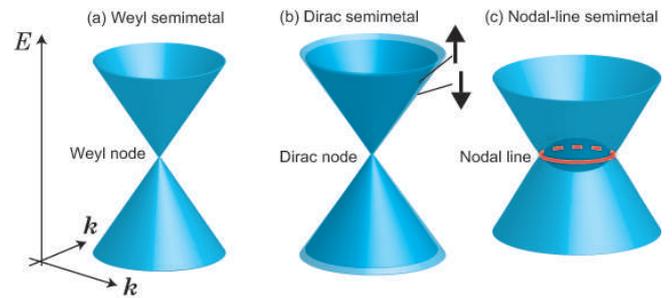}
\caption{(Color online) Examples of topological semimetals (topological metals): (a) Weyl semimetal, (b) Dirac semimetal, and 
(c) nodal-line semimetal. In (a) the Dirac cone is not degenerate and in (b) it is 
Kramers degenerate.}
\label{fig:topologicalSM}
\end{center}
\end{figure}

In Fig.~\ref{fig:topologicalSM}, we show typical classes
of three-dimensional (3D) topological semimetals.
In Weyl semimetals (WSMs)\cite{Murakami07b,Wan12} [Fig.~\ref{fig:topologicalSM}(a)], the bulk valence and conduction 
bands touch each other at isolated points called Weyl nodes, and around the Weyl nodes the bands form 
nondegenerate 3D Dirac cones. 
In spinful systems, i.e., when the spin-orbit coupling is nonzero, if both time-reversal (TR)
and inversion symmetries are present,  all states are doubly degenerate by the Kramers theorem. Therefore, a Weyl semimetal 
requires breaking of either inversion symmetry or TR symmetry. 
On the other hand, a topological semimetal having a Dirac cone with Kramers double degeneracy 
is realized when both inversion symmetry and TR symmetry are preserved.
Such a semimetal is called a Dirac semimetal \cite{Wang12,Wang13} (Fig.~\ref{fig:topologicalSM}(b)). 
Another type of topological metal called nodal-line semimetals (NLSs) \cite{Mullen15,Fang15,Chen15nano,Weng15b,Kim15,Yu15,Xie15,Chan15,Zeng15,Yamakage16,Zhao15}.
is shown in Fig.~\ref{fig:topologicalSM}(c), where the band gap closes along a loop in ${\bf k}$ space. 
In these topological semimetals, the band gap closes at generic points in ${\bf k}$ space, which originates
from the interplay between the ${\bf k}$-space topology and symmetry.
Because degeneracies in topological semimetals are
not necessarily located at high-symmetry points  in the Brillouin zone, an
efficient and systematic search for  topological semimetals is usually difficult. 
Moreover, such degeneracies occur at generic ${\bf k}$ points;
therefore, they may easily be overlooked in {\it ab initio} calculations, where 
the band structure is usually calculated only along high-symmetry lines.
Thus, the search for topological semimetals has been an elusive issue.

In this paper, we first review basic properties of topological semimetals, and then review the {\it ab initio} approach to the material search and explorations of physical properties of topological semimetals. In Sect.~2, we review basic properties of WSMs and 
show some {\it ab initio} results for WSMs. 
We then give an overview of Dirac semimetals in Sect.~3.
In Sect.~4, we explain the NLSs and give some examples such as calcium.
Numerical methods for topological systems in {\it ab initio} calculations are 
explained in Sect.~5.
We conclude this paper in Sect.~6.

\section{Weyl Semimetals}
\subsection{Overview}
\begin{figure}[h]
\begin{center}
\includegraphics[width=85mm]{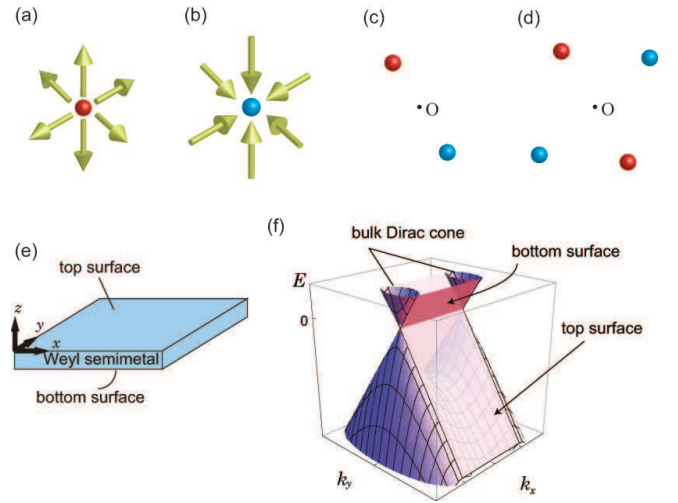}
\caption{(Color online) Topological characterization of Weyl nodes. (a) Monopole and (b) antimonopole
having the monopole charge $+1$ and $-1$, respectively. (c) In systems with inversion symmetry but without TR symmetry, the monopole
charges of the Weyl nodes are antisymmetric in ${\bf k}$ space. (d) In systems with TR symmetry but 
 without inversion symmetry, the monopole
charges of the Weyl nodes are symmetric in ${\bf k}$ space.
(e) Slab of a Weyl semimetal. At both surfaces of the slab,  Fermi arc surface states appear. (f) Dispersion of the bulk states and the surface Fermi arc. Between the bulk Dirac cones, Fermi arcs appear on the respective surfaces of the slab. 
}
\label{fig:Weyl}
\end{center}
\end{figure}

In a WSM, \cite{Murakami07b,Wan,Yang,Witczak-Krempa} 
the  Weyl nodes have a topological nature.
Namely, one can associate each Weyl node with
a number $\pm 1$, called a monopole charge for the Berry curvature in ${\bf k}$ space\cite{Klinkhamer,Murakami07b,Jiang}. 
The monopole charge is also called the helicity or chirality.
This means that the degeneracy at a Weyl node cannot be lifted perturbatively. 
The monopole density $\rho^{(n)}({\bf k})$, associated with the Berry curvature $\Omega^{(n)}({\bf k})$, is defined as 
\begin{equation}
\Omega^{(n)}({\bf k})=i\left\langle \frac{\partial u_{n{\bf k}}}{\partial {\bf k}}\right|
\times\left| \frac{\partial u_{n{\bf k}}}{\partial {\bf k}}\right\rangle,\ \ 
\rho^{(n)}({\bf k})=\frac{1}{2\pi}
\nabla_{\bf k}\cdot\Omega^{(n)}({\bf k}),
\end{equation}
where $|u_{n{\bf k}}\rangle$ is the periodic part of the $n$th Bloch eigenfunction.
One can prove that in general the monopole density is a superposition of $\delta$ functions with integer coefficients, 
$\rho^{(n)}({\bf k})=\sum_l q_l\delta({\bf k}-{\bf k}_l)$ ($q_l$: integer), and this $q_l$ is the monopole charge at ${\bf k}_l$. The positions of the monopoles ${\bf k}={\bf k}_l$ are the ${\bf k}$ points
where the $n$th band touches other bands in terms of energy. 
In particular, the monopole charge $q_l$ at a Weyl node is either $+1$ or $-1$, corresponding to a monopole (Fig.~\ref{fig:Weyl}(a)) or an antimonopole (Fig.~\ref{fig:Weyl}(b)), respectively.
This is seen directly by using 
a simple 2-band Hamiltonian representing a Weyl node at ${\bf k}_0$, 
$H=\sum_{i,j}b_{ij}\sigma_i (k_j-k_{0j})$, where $b_{ij}$ are constants and $\sigma_i$ are the Pauli matrices; then, the monopole charges at the Weyl node for the upper band and  lower band are
calculated as $q^{(\rm upper)}=-q^{(\rm lower)}
=-{\rm sgn}({\rm det} b_{ij})(=\pm 1)$, where ${\rm det} b_{ij}$ is the determinant of the 3$\times $3 matrix of $b_{ij}$. 
The Weyl nodes in three dimensions are topological and robust against perturbations owing to their 
monopole charges, unlike those in two dimensions. 

We explain here the symmetry requirements of WSMs. 
WSMs appear only when either the TR symmetry or the inversion symmetry is 
broken. 
In WSMs with inversion symmetry but without TR symmetry, 
the monopole density is an odd function of ${\bf k}$, and the Weyl nodes distribute antisymmetrically in ${\bf k}$ space
(Fig.~\ref{fig:Weyl}(c)). 
Similarly, in WSMs with TR symmetry but without inversion symmetry,  the monopole density is an even function of ${\bf k}$, and 
the Weyl nodes distribute symmetrically in ${\bf k}$ space.
Since the sum of the monopole charges over the entire Brillouin zone is zero, the minimal configuration of 
Weyl nodes in this case contains four Weyl nodes, as shown in Fig.~\ref{fig:Weyl}(d).

This topological property of Weyl nodes \cite{Berry84, Volovik, Murakami07b} necessitates the appearance of a surface Fermi surface 
forming an open arc, called a Fermi arc 
\cite{Wan12,Yang,Witczak-Krempa, Ojanen, Okugawa, Haldane}, unlike usual Fermi surfaces, which are closed surfaces (in three dimensions) or closed loops (in two dimensions).
The two end points of the Fermi arc are projections of the Weyl points, a monopole and an antimonopole, onto the surface Brillouin zone. 
The appearance of a surface  Fermi arc is shown by introducing the Chern number in 
a two-dimensional slice of the three-dimensional Brillouin zone \cite{Wan12}. 
In a slab of a WSM (Fig.~\ref{fig:Weyl}(e)), a typical form of the dispersion of bulk and surface states for the surface along the $xy$ plane is shown 
in Fig.~\ref{fig:Weyl}(f). Between the two bulk Dirac cones, there are two surface Fermi arcs, 
one on the top surface  and the other on the bottom surface. These two Fermi arcs are both tangential to 
the bulk Dirac cones but have opposite dispersions and thus opposite velocities. 

WSMs with broken TR symmetry have been proposed to include pyrochlore iridates \cite{Wan12}, HgCr$_2$Se$_4$ \cite{Xu11},  
Co-based Heusler compounds \cite{Wang16},
magnetic superlattices with a TI and a normal insulator (NI) \cite{BurkovPRL,BurkovPRB}, and so on. 
Those with broken inversion symmetry include TaAs, BiTeI under high pressure\cite{Liu-BiTeI},
Te at high pressure\cite{Hirayama15}, 
LaBi$_{1-x}$Sb$_x$Te$_3$, LuBi$_{1-x}$Sb$_x$Te$_3$ \cite{Liu-BiTeI},  
transition-metal dichalcogenides \cite{Sun15,Weng15}, 
SrSi$_2$ \cite{Huang-SrSi2}, WTe$_2$ \cite{Soluyanov-WTe2}, 
HgTe$_{x}$S$_{1-x}$ under strain \cite{Rauch},
and superlattice with broken inversion symmetry \cite{Halasz}. In accordance with theoretical predictions \cite{Weng15,Huang},  
TaAs and related materials have been experimentally established to 
be WSMs \cite{Lv15,Xu15b,Lv15b,Xu15c,Yang15}.

Weyl nodes are generally difficult to find in the 3D Brillouin zone because they 
are at general positions in $\bm{k}$ space. 
Meanwhile, when the system breaks TR symmetry but preserves inversion symmetry,
we can exploit a $\mathbb{Z}_2$ topological invariant to find a WSM phase \cite{Hughes, Turner}. 
Let $\bm{K}_i$ denote wavevectors satisfying $\bm{K}_{i}\equiv -\bm{K}_{i}$
modulo reciprocal vectors. They are called time-reversal invariant momenta (TRIM).
The $\mathbb{Z}_2$ invariant is given by
\begin{align}
\chi _P=\prod _{\bm{K}_i:\ {\rm TRIM}}\prod _{n}^{{\rm occ.}}\xi _n(\bm{K}_i),
\label{eq:chiP}
\end{align}
where $\xi _n(\bm{K}_i)\ (=\pm 1)$ is the parity eigenvalue of the $n$th occupied band at one
ot the TRIM
$\bm{K}_i$.
The topological invariant $\chi _P$ takes a value of $1$ or $-1$.
If $\chi _P=-1$,
the energy bands have $(2N+1)$ pairs of Weyl nodes, where $N$ is an integer \cite{Hughes, Turner}. 
For example, ZrCo$_2$Sn, which is a candidate magnetic WSM,
has a nontrivial $\mathbb{Z}_2$ invariant \cite{Wang16}.
However, this $\mathbb{Z}_2$ invariant cannot be defined in a system without inversion symmetry, leading 
to a further difficulty in determining whether the system is a WSM
since Weyl nodes are generally difficult to find in the 3D Brillouin zone.
In such inversion-asymmetric WSMs, 
the trajectories of the emergent Weyl nodes upon a continuous change in the system 
are related to
the Fu--Kane--Mele $\mathbb{Z}_2$ invariant $\nu _0$ \cite{Fu07, Murakami07b}, whose definition is similar to that of $\chi_P$, as discussed later in this section.
Studies on this relationship provide us with a new perspective on the search for semimetals through {\it ab initio} calculations.

\subsection{Phase transition between topological insulators and Weyl semimetals}
\label{sec:phasetransition}
When the inversion symmetry is broken, there is a close relationship between 
TIs and WSMs in three dimensions. Namely, in 3D inversion-asymmetric systems, phase transitions between two phases with a different 
$\mathbb{Z}_2$ topological number $\nu_0$ are always accompanied by the 
WSM phase in between \cite{Murakami07b,MurakamiKuga}.
We assume that the TR symmetry is preserved. 
Figure \ref{fig:Weyl-phasediagram}(a) shows a universal phase diagram for STI-WTI or STI-NI phase transitions, where $m$ is a parameter controlling the phase transition 
and $\delta$ is a parameter controlling the degree of inversion symmetry breaking. Here, STI and WTI stand for a strong TI and a weak TI, 
respectively.  
As seen from Fig.~\ref{fig:Weyl-phasediagram}(a), the phase transition is different in the presence or absence of inversion 
symmetry.
First, in inversion-symmetric systems, 
the gap can close only at TRIM,
where the parity eigenvalues are exchanged at the gap closing (Fig.~\ref{fig:Weyl-phasediagram}(b)); thereby, the $\mathbb{Z}_2$ topological numbers change, 
leading to a STI-WTI or STI-NI transition. At the band touching, the system is a Dirac semimetal (DSM).
Such a phase transition is realized, for example, in TlBi(S$_{1-x}$Se$_2$)$_2$ 
at around $x\sim 0.5$ \cite{Xu-TlBiSSe,Sato-TlBiSSe}. 
Second, in inversion asymmetric cases, every state is nondegenerate except at TRIM. 
The gap can close only away from TRIM. 
Such gap closing 
always involves the creation of a pair of Weyl nodes \cite{Murakami07b,MurakamiKuga}.  Because the Weyl nodes are either monopoles or antimonopoles, 
they can appear or disappear by pair creation/annihilation only. 
This gives robustness to the WSM phase; namely, as we change an external parameter $m$ which controls the topological phase transition, 
the WSM phase persists for a finite range of $m$: $m_1<m<m_2$. At the end value $m=m_1$ 
($m_2$), the pair creation (annihilation) of Weyl nodes occurs. These considerations lead to the phase diagram shown in 
Fig.~\ref{fig:Weyl-phasediagram}(a).

The pattern of monopole-antimonopole pair creation and annihilation depends on the
crystallographic symmetry. From Fig.~\ref{fig:Weyl}(d), the minimal case involves two monopoles and two antimonopoles, as shown in  Fig.~\ref{fig:Weyl-phasediagram}(c). As we change the  control parameter $m$ from the value for an insulating phase, 
two monopole-antimonopole pairs are created at first. As we change $m$ further, these monopoles and antimonopoles
move in ${\bf k}$ space, and eventually they annihilate pairwise. 
When the system has discrete rotational symmetry such as threefold or fourfold rotational symmetry, the minimal number of monopole-antimonopole pairs will 
be larger, as demonstrated in some systems \cite{Halasz}. 
In this phase transition from the WSM phase to the TI phase, the Fermi arcs in the WSM phase 
merge with each other to form a Dirac cone in the TI phase \cite{Okugawa}. 
Here we emphasize the role of topology. The appearance of the WSM phase in three dimensions in a finite range
in the phase diagram is due to the topological nature of Weyl nodes \cite{MurakamiKuga}.  In contrast, in two  dimensions, 
the NI-TI phase transition occurs directly, 
without the appearance of a WSM phase \cite{Murakami07b,MurakamiKuga}. 
This result has been shown by assuming no additional crystallographic symmetries.
Furthermore, the result in Ref.~\citen{Murakami-gapclosing} shows that this conclusion of the existence of the WSM phase in the $\mathbb{Z}_2$ phase transition
holds true in general, whenever inversion symmetry is broken. 
One of the remarkable 
conclusions here is that an insulator-to-insulator (ITI) transition never occurs in inversion-asymmetric systems. 
This is in strong contrast with inversion-symmetric systems,
where a transition between different $\mathbb{Z}_2$ topological 
phases always occurs as an ITI transition (i.e., at a single value of $m$).

Indeed, this NI-WSM-STI phase transition appears in various materials without 
inversion symmetry. 
We first apply this theory to BiTeI, whose space group is No.156 ($P3m1$) and lacks inversion symmetry \cite{ishizaka11}. 
BiTeI is an NI at ambient pressure and has been proposed to 
become an STI at high pressure \cite{Bahramy,Yang-BiTeI}. 
Subsequently, {\it ab initio} calculations \cite{Liu-BiTeI} showed the existence of a WSM phase between 
the NI and STI phases in a narrow window of pressure, 
which had been overlooked in earlier works
\cite{Bahramy,Yang-BiTeI}.
When the pressure is increased, the gap first closes at six S points on the A-H lines, 
and six pairs of Weyl nodes are created. Then the Weyl nodes move in ${\bf k}$ space until 
they annihilate each other on mirror planes, leading to the STI phase \cite{Liu-BiTeI}. 
The combination of all the trajectories of the Weyl nodes forms a loop around the TRIM point (A point), meaning that a band inversion occurs at the A point. This leads to a change in the $\mathbb{Z}_2$ topological number between the two bulk-insulating phases 
sandwiching the WSM phase \cite{Murakami07b,MurakamiKuga}.

Other examples are LaBi$_{1-x}$Sb$_x$Te$_3$ and LuBi$_{1-x}$Sb$_x$Te$_3$, which 
have been proposed to show NI-WSM -TI phase transitions
upon changing $x$ \cite{Liu-BiTeI}. 
The space group is No.160 ($R3m$) and lacks inversion symmetry. 
By increasing $x$, the band gap closes at
six generic points, and Weyl nodes are created in pairs \cite{Liu-BiTeI}. 
The resulting 12 Weyl nodes move in ${\bf k}$ space, and they are then annihilated pairwise
to open a bulk gap, and the $\mathbb{Z}_2$ topological number is changed from that in the low-$x$ phase \cite{Liu-BiTeI}. 

\subsection{Closing of a gap in inversion-asymmetric systems}
\label{sec:closing}
\begin{figure}[htp]
\begin{center}
\includegraphics[width=8cm]{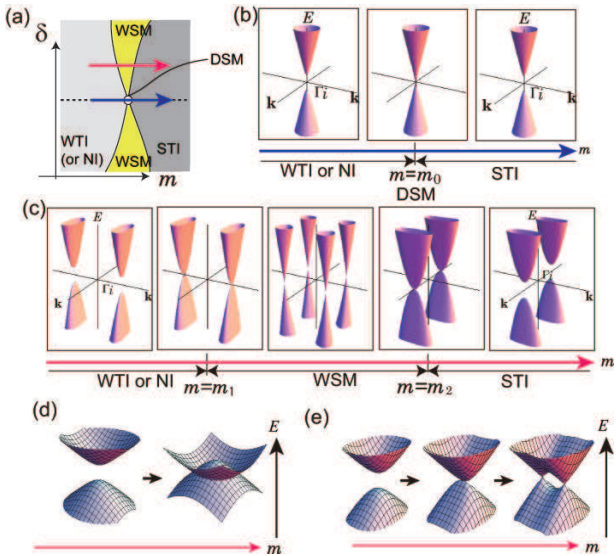}
\caption{(Color online)  (a) Universal phase diagram for $\mathbb{Z}_2$ topological phase transitions shown as a function of an external parameter $m$ and a parameter $\delta$ controlling the degree of inversion symmetry breaking \cite{Murakami07b}. The horizontal dotted line 
is the inversion-symmetric line. (b) (c) Evolutions of bulk band structure with a change in
the parameter $m$ for inversion-symmetric (b) and inversion-asymmetric (c)
cases. The blue and red arrows in (a) correspond to the cases in (b) and (c), respectively.
(d) (e) Band evolutions towards gap closing. 
(d) and (e) show schematic illustrations for two classes of band evolution after gap closing
in semiconductors without inversion symmetry. 
In (d) the system becomes an NLS, and in (e) the system becomes a WSM.
(From Sci. Adv. 3, e1602680 (2017) \cite{Murakami-gapclosing}. Reprinted with permission from AAAS.)}
\label{fig:Weyl-phasediagram}
\end{center}

\end{figure}

Next, motivated by 
the universal phase diagram (Fig.~\ref{fig:Weyl-phasediagram}(a)), we can show a stronger conclusion that
the closing of a gap of any inversion-asymmetric semiconductor (with spin-orbit coupling
and TR symmetry) in three dimensions leads to either a WSM or an NLS 
\cite{Murakami-gapclosing}, as we explain in the following.
This is shown for all the space groups without inversion symmetry, and is widely applicable to
various semiconductors without inversion symmetry.

To show this, we introduce a single parameter $m$ in the Hamiltonian, and assume that 
the 
change in $m$ does not change the system symmetry.
We consider a Hamiltonian matrix  $H({\bf k},m)$ representing the 
lowest conduction band and the highest valence band. Here, ${\bf k}$ is
a Bloch wavevector. 
We assume that for $m<m_0$ the 
system is an insulator, and that at $m=m_0$ the gap closes at a wavevector ${\bf k}={\bf k}_0$.
Then to see what happens when $m$ exceeds $m_0$, we expand the Hamiltonian in terms of $m-m_0$ and ${\bf q}\equiv{\bf k}-{\bf k}_0$, and 
retain some of the lower-order terms. We consider 
all 138 space groups without inversion symmetry. 
For each space group, there are various high-symmetry 
${\bf k}$ points such as $\Gamma$, $X$, and $L$.
Each ${\bf k}$ point is associated with a little group ($k$ group), which leaves the ${\bf k}$ point unchanged. 
Then the Hamiltonian is determined by the double-valued irreducible representations (irreps) \cite{Bradley} of the  little group at ${\bf k}_0$. 
Let $R_{\rm c}$ and $R_{\rm v}$ denote the irreps of the lowest conduction band and the highest valence band, respectively.

\begin{figure}[htp]
\begin{center}
\includegraphics[width=8.5cm]{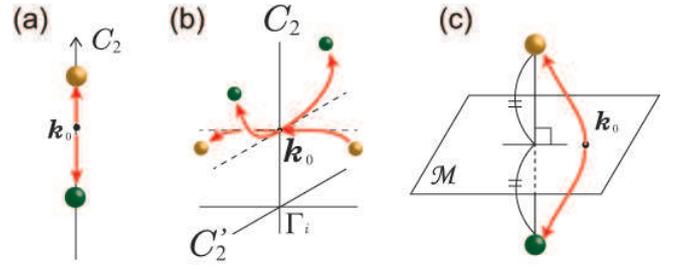}
\caption{(Color online) Trajectories of Weyl nodes at closing of the band gap in inversion-asymmetric insulators.
(a) Case with a little group having $C_2$ symmetry and the two bands having
different $C_2$ eigenvalues. 
(b) Case with a little group having $C_2$ and $\Theta C'_2$ symmetries and the two bands having the
same $C_2$ eigenvalues. 
(c) 
Case with a little group having mirror symmetry and the two bands having the
same mirror eigenvalues.  Yellow and green spheres denote
monopoles and antimonopoles in ${\bf k}$ space, respectively, and they are both Weyl nodes. 
(From Sci. Adv. 3, e1602680 (2017) \cite{Murakami-gapclosing}. Reprinted with permission from AAAS.)}
\label{fig:trajectory}
\end{center}
\end{figure}

As the first example, we consider the case when the band gap closes at generic ${\bf k}$ points in three
 dimensions. Such closing of the gap 
always accompanies the creation of a pair of Weyl nodes, as shown in Refs.~\citen{Murakami07b} 
and \citen{MurakamiKuga}. 
This occurs because Weyl nodes have quantized monopole charges $q=\pm 1$ for the
${\bf k}$-space Berry curvature; this topological property allows the creation of pairs of Weyl nodes 
with $q=+1$ and $q=-1$. Furthermore, if there is additional symmetry, the gap closing becomes 
more involved. For example, suppose the little group 
consists only of the twofold ($C_2$) symmetry. 
Then there are two one-dimensional (1D) 
irreps of the $C_2$ eigenvalues with opposite signs. 
For $R_{\rm c}\neq R_{\rm v}$, the gap can close and the closing of the gap 
accompanies the creation of a pair of Weyl nodes. 
When $m$ is increased, the two Weyl nodes (a monopole and an antimonopole) move along the 
$C_2$ axis, as shown in Fig.~\ref{fig:trajectory}(a).
On the other hand, when $R_{\rm c}=R_{\rm v}$, the gap cannot close by changing $m$ because of level repulsion. In addition, when there are some additional symmetries such as
$\Theta C'_2$, where $\Theta$ stands for the TR operation and $C'_2$ represents 
another twofold rotation, the gap can close when $R_{\rm c}=R_{\rm v}$ and the trajectory of the Weyl nodes is shown in 
Fig.~\ref{fig:trajectory}(b).
Next we consider the case with the little group
consisting only of a mirror symmetry or a glide symmetry. 
There are two representations with opposite signs of mirror (or glide) eigenvalues. 
In the case $R_{\rm c}=R_{\rm v}$, the gap can close, leading to 
the creation of a pair of Weyl nodes. The trajectory of the monopole and that of 
the antimonopole are mirror images to each other (Fig.~\ref{fig:trajectory}(c)). 
On the other hand, for $R_{\rm c}\neq R_{\rm v}$, 
the gap closes along a loop (i.e., nodal line) within the mirror plane in ${\bf k}$ space, and the
system is an NLS. 

From this analysis, we conclude that there are only two possibilities after closing of the gap 
in inversion-asymmetric insulators with TR symmetry: NLSs (Fig.~\ref{fig:Weyl-phasediagram}(d)) and 
WSMs  (Fig.~\ref{fig:Weyl-phasediagram}(e)). 
The space group, the wavevector at the gap closing, and the irreducible representations
of the highest valence and lowest conduction bands uniquely 
determine which possibility is realized and where the gap-closing points or lines are located
after the closing of the gap, as summarized in Ref.~\citen{Murakami-gapclosing}.
In the NLSs, the highest valence and lowest conduction 
bands have different mirror eigenvalues.
Remarkably, an ITI 
phase transition never occurs in any inversion asymmetric system, as we noted earlier.

\subsection{Realization of materials}
This universal result applies to all crystalline materials with spin-orbit coupling (SOC) 
and TR symmetry without inversion symmetry. We present some examples here.
The first example is HgTe$_{x}$S$_{1-x}$ under strain, which has been
shown to become a WSM \cite{Rauch}.
It has a zincblende structure having the space group $F\bar{4} 3 m$ (No.216), and a 
strain along the [001] direction reduces the space group to $I\bar{4}m2$ (No.119).
In this case, when $x$ is increased, the gap closes at four points on the $\Gamma$-K lines, producing four pairs of Weyl nodes. The eight Weyl points then move within the $k_z=0$ plane with a further increase in $x$, until they are
mutually annihilated after $\pm\pi/2$ rotation around the 
$[001]$ axis \cite{Rauch}. 

\begin{figure}[htp]
\begin{center}
\includegraphics[width=8cm]{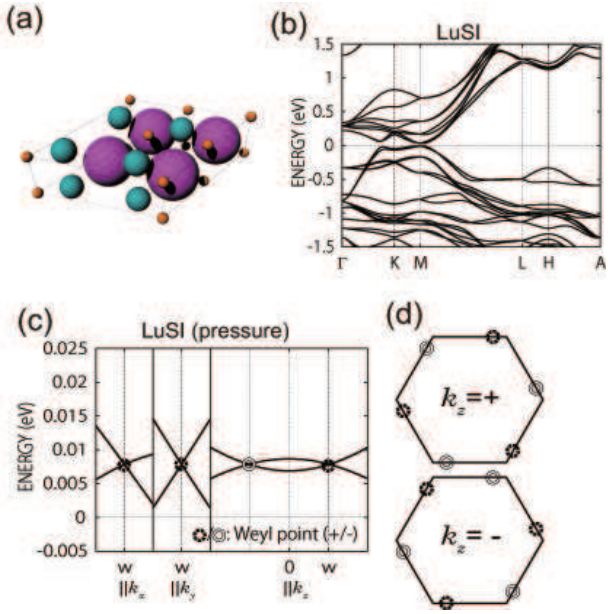}
\caption{(Color online) Emergence of a WSM phase in LuSI under pressure.
(a) Crystal structure of LuSI. 
(b) Electronic band structure of LuSI.
(c) Band structure of 
LuSI under pressure, where the lattice constant $c$ is $0.945$ times the
value at ambient pressure.
(d) Positions of the Weyl nodes of LuSI under pressure on the $k_z = 0$ plane.
The energy is measured from the Fermi level. (From Sci. Adv. 3, e1602680 (2017)  \cite{Murakami-gapclosing}. Reprinted with permission from AAAS.)}
\label{fig:LuSI}
\end{center}
\end{figure}

We can use the result in the previous subsection 
to search for topological semimetal materials.
We start with any narrow-gap semiconductor without inversion symmetry, and 
close the gap by changing the system by, for example, applying pressure or by atomic substitution. Then, as we showed 
previously, the system becomes either a WSM or an NLS.
As an example, we show  LuSI as a WSM under pressure.
The structure type of LuSI is GdSI (Fig.~\ref{fig:LuSI}(a)) with the space group No.174 ($P\bar{6}$). 
Figure \ref{fig:LuSI}(b) shows the electronic structure of LuSI, having a 
very narrow gap ($< 0.04$ eV) near the M point.
By applying pressure, the band gap first closes at six generic points on the $k_z=0$ plane.
The system then becomes a WSM under pressure, having
six monopoles and six antimonopoles at the same energy since they are related by symmetry, 
as shown in 
Fig.~\ref{fig:LuSI}(c)).

\begin{figure}[ptb]
\begin{center}
\includegraphics[clip,width=8cm]{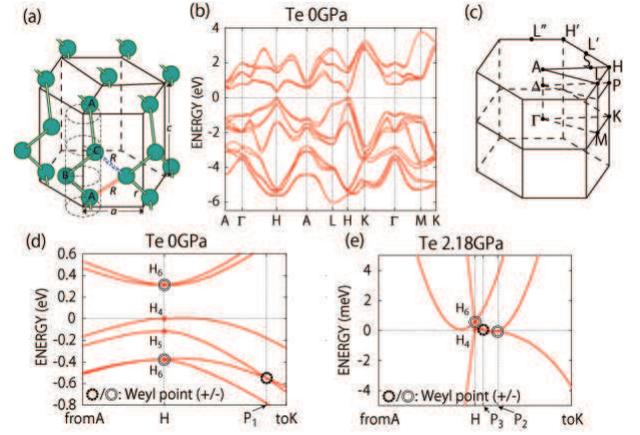} 
\caption{\label{bndTeSe}(Color online) 
(a) Crystal structure and (b) electronic structures of Te at ambient pressure obtained by the $GW$+SO.
(c) High-symmetry points in the Brillouin zone.
(d) Magnified figure of electronic structure for Te.
(e) Electronic structures of Te at $2.18$ GPa.
P$_2$ and P$_3$ are Weyl nodes on the H-A (H-K) line.
Weyl points are indicated by the 
black dashed/white solid circles having a positive/negative monopole charge
(calculated from the lower band around the degenerate point).
The energy is measured from the Fermi level.
(From Sci. Adv. 3, e1602680 (2017) \cite{Murakami-gapclosing}. Reprinted with permission from AAAS.)}
\end{center}
\end{figure}

Another example is 
 tellurium (Te) \cite{Hirayama15}. 
The crystal structure consists of 1D helical chains, as shown in Fig.~\ref{bndTeSe}(a).
The unit cell contains three atoms in the same helical chain. 
When the helical chains are right-handed/left-handed, its space group is No.152 ($P3_121$)/No.154  ($P3_221$), and these two
structures are mirror images of 
each other. 
Tellurium is a narrow gap semiconductor at ambient pressure.
Figure~\ref{bndTeSe}(b) shows the band structures of Te at ambient pressure, and 
Fig.~\ref{bndTeSe}(c) shows the labels for high-symmetry points.
The band gap in the $GW$+SO is $0.314$ eV for Te, 
which is in good agreement with the experimental value of $0.323$ eV~\cite{anzin77,tutihasi67}.
Both the bottom of the conduction band and top of the valence band 
are close to but slightly away from the H point in Te (Fig.~\ref{bndTeSe}(c)) \cite{doi70}. 
At higher pressure, the gap closes and eventually a pair of Weyl nodes
is produced at each of the four P points on the K-H lines, 
and the system becomes a WSM at 2.17--2.19 GPa (Fig.~\ref{bndTeSe}(e)).
The P$_2$ and P$_3$ points are Weyl nodes between the valence and conduction bands.
The Weyl nodes then move along
the  $C_3$ axes (K-H lines),
as was confirmed by the 
{\it ab initio} calculation \cite{Hirayama15}.



%

\section{Dirac Semimetals}
Materials having a linear dispersion with Kramers degeneracy at or near the Fermi energy are called Dirac semimetals.
Thus, from the requirement for the Kramers degeneracy, inversion and TR symmetries are required for Dirac semimetals.
Dirac semimetals have been proposed to include Na$_3$Bi \cite{Wang12}, Cd$_3$As$_2$ \cite{Wang13},  $\beta$-BiO$_2$ \cite{Young12}, and so on, 
and they have been experimentally established in Na$_3$Bi \cite{Liu14s,Xu15} and Cd$_3$As$_2$ \cite{Neupane14,Borisenko14}.
The Dirac nodes can be regarded as a superposition of two Weyl nodes 
with monopole charges $q=\pm 1$, i.e., a monopole and an antimonopole
for the Berry curvature.
Thus, by adding perturbation 
the gap can be opened \cite{Young12} via monopole-antimonopole 
pair annihilation. 

Thus, in real materials, one can always have such perturbation terms. Dirac nodes always 
have a gap if no other crystallographic symmetries are considered, 
and are not robust against perturbations. This is in strong contrast to Weyl nodes. 
The crystallographic symmetries required for stabilization of 
the Dirac semimetal phase have been discussed \cite{Yang-DSM}, and indeed such symmetries
are present in the material examples mentioned above. 

\section{Nodal-Line Semimetal}
As another example of topological semimetals, in an NLS \cite{BurkovPRB,Mullen15,Fang15,Chen15nano,Weng15b,Kim15,
Yu15,Xie15,Chan15,Zeng15,Yamakage16,Zhao15,Hirayama-Ca}, 
the band gap closes along a curve, called a nodal line, in ${\bf k}$ space. Such a degeneracy along a nodal line
cannot appear without a special reason, such as the symmetry or topology of the system.
There are several mechanisms
for the emergence of nodal lines. 
Among the various mechanisms of the emergence of nodal lines, we introduce two typical mechanisms: (A) mirror symmetry and (B) the $\pi$ Berry phase. In both cases, the nodal line forms a closed loop. 
In the following, we discuss the two cases (A) and (B) separately.
 In some materials the two mechanisms coexist, whereas in others 
the nodal line originates from only one of these mechanisms. 

\subsection{Nodal lines stemming from mirror symmetry}
In case (A), the system should have a mirror symmetry or glide symmetry.
When the mirror/glide operation and the Hamiltonian commute, they are 
simultaneously diagonalized. The eigenvalues of the mirror/glide operations 
depend on the situation, but they always have 
two values with opposite signs. For example, in spinless systems, the mirror eigenvalues are
${\cal M}=\pm 1$, originating from ${\cal M}^2=1$, and in spinful electronic systems, 
the mirror eigenvalues are ${\cal M}=\pm i$.
In all cases, the states on the mirror plane can be classified into two classes based on 
the mirror/glide eigenvalues.
If the valence and conduction bands have different mirror/glide eigenvalues, then the two bands
have no hybridization (i.e., no level repulsion) on the mirror/glide plane, even if the two bands approach and cross. 
This results in a degeneracy along a loop on the mirror/glide plane. 
This line-node degeneracy is 
protected by mirror/glide symmetry; once the mirror/glide symmetry is broken, the degeneracy is
lifted in general.

Many materials have been proposed to belong to this class of NLSs.
 Dirac NLSs having Kramers degeneracy include
carbon allotropes \cite{Chen15nano,Weng15b}, 
Cu$_3$PdN \cite{Kim15,Yu15}, Ca$_3$P$_2$ \cite{Xie15,Chan15},
LaN \cite{Zeng15}, CaAgX (X=P, As) \cite{Yamakage16} and 
compressed black phosphorus \cite{Zhao15}. Weyl NLSs
with no Kramers degeneracy
 include HgCr$_2$Se$_4$ \cite{Xu11} and TlTaSe$_2$ \cite{Bian15}.

\begin{figure}[htp]
\includegraphics[width=8cm]{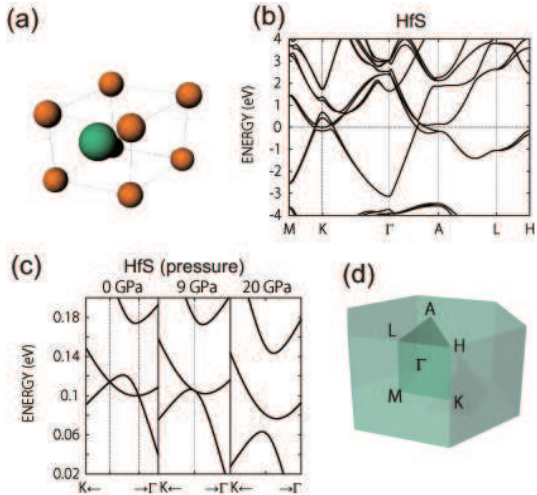}
\caption{(Color online) NLS phase in HfS. 
 (a) Crystal structure of HfS. (b) Electronic band structure of Hf at ambient pressure.
(c) Phase transition of HfS under pressure.
The energy is measured from the Fermi level.
(d) Brillouin zone of HfS. (From Sci. Adv. 3, e1602680 (2017) \cite{Murakami-gapclosing}. Reprinted with permission from AAAS.)
}
\label{fig:HfS}
\end{figure}

Examples of NLSs can be found by the argument in 
Sect.~\ref{sec:closing}. One can start with a narrow-gap semiconductor and close the gap on 
a mirror/glide plane, and one can obtain an NLS when the mirror/glide
eigenvalues are different between the two bands involved. 
For example, we find that HfS has nodal lines near the Fermi level.
The structure of HfS is shown in Fig.~\ref{fig:HfS}(a), which has the space group No.187 ($P\bar{6}m2$).
Figure~\ref{fig:HfS}(b) shows the electronic band structure of HfS.
If the SOC is neglected, Dirac nodal lines exist around the K points on the $k_z =0$ plane.
The SOC lifts the degeneracy of the Dirac nodal lines, and Weyl nodal lines appear instead near the K points on the mirror plane 
$k_z =0$.
By applying pressure (Fig.~\ref{fig:HfS}(c)) or by atomic substitution from S to Se, the nodal lines become smaller.
At 9 GPa, the nodal lines shrink to points, and 
then a gap opens above 9 GPa.

\subsection{Nodal lines stemming from the $\pi$ Berry phase}
\label{sec:nodalline}
\begin{figure}[h]
\begin{center}
\includegraphics[width=75mm]{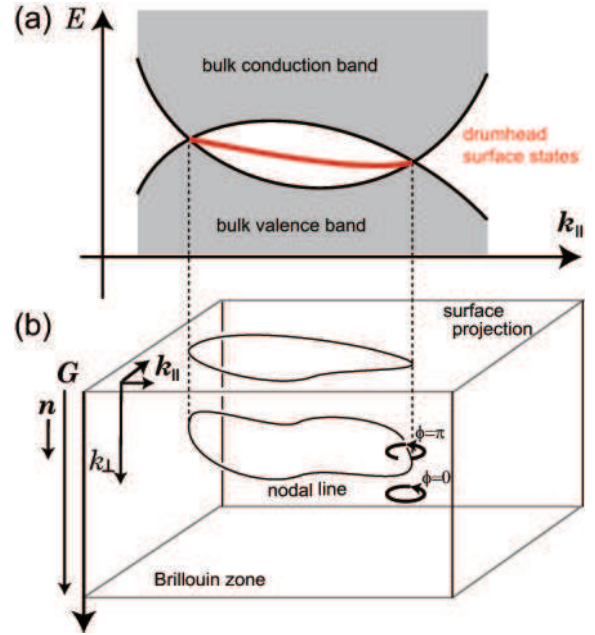}
\caption{(Color online) (a) Schematic dispersion for surface and bulk states of an NLS. 
Drumhead surface states (red), which appear within the region surrounded
by the surface projection of the nodal line.
(b) Corresponding nodal line in the 3D Brillouin zone. 
The Berry phase along a loop encircling the nodal line is $\pi$.
 }
\label{fig:nodalline}
\end{center}
\end{figure}

The second mechanism (B) occurs in spinless systems with inversion and TR
symmetries. Spinless systems include electronic systems with zero SOC and 
bosonic systems such as magnonic and photonic systems. 
Examples of electronic systems include calcium at high pressure \cite{Hirayama-Ca}
and AX$_2$ (A = Ca, Sr, Ba; X = Si, Ge, Sn) \cite{Huang-Liu}.
The Berry phase
around the nodal line is $\pi$, and this $\pi$ Berry phase
topologically protects the nodal line.
In spinless systems with inversion and TR symmetries, the Berry
phase along any closed loop is quantized as an integer multiple of $\pi$, as 
shown in the following. 
Here, the Berry phase $\phi (\ell)$ along a loop $\ell$ 
is defined as
\begin{align}
\phi (\ell)=-i\sum_{n}^{\text{occ.}}\int_{\ell} d {\bf k}\cdot 
\left\langle{u_n({\bf k})}\right| \nabla_{\bf k} \left|{u_n({\bf k})}\right\rangle,
\label{eq:phi}
\end{align}
where
the sum is over the occupied states and the system is assumed to have a gap
everywhere along loop $\ell$.
The Berry phase is defined modulo $2\pi$, corresponding to a gauge 
degree of freedom. Under the product of TR and spatial inversion operations, 
this quantity is transformed into $-\phi(\ell)$, and therefore we obtain
$\phi(\ell)\equiv -\phi(\ell) \ \ ({\rm mod}\ 2\pi)$, i.e., 
$\phi(\ell)\equiv 0\ {\rm or}\ \pi\  ({\rm mod}\ 2\pi)$ for any loop $\ell$.
The Berry phase around a nodal line from this mechanism (B) is $\pi$ (Fig.~\ref{fig:nodalline}(b)), and 
therefore the nodal line is protected because of the quantization of the Berry phase. 

Because of the topological character of these nodal lines,
new $\mathbb{Z}_2$ topological invariants $\nu _i$ ($i=0,1,2,3$) can be introduced to identify NLSs \cite{Kim15}.
The $\mathbb{Z}_2$ invariants are 
\begin{align}
(-1)^{\nu _0}=\prod _{n_j=0,1}\prod _m^{\rm occ.}\xi _m(\bm{\Gamma}_{n_1n_2n_3}), \\
(-1)^{\nu _i }=\prod _{n_i=1,n_{j\neq i}=0,1}\prod _m^{\rm occ.}\xi _m(\bm{\Gamma}_{n_1n_2n_3}),
\end{align}
where $\xi _m(\bm{\Gamma}_{n_1n_2n_3})$ is the parity eigenvalue of the $m$th occupied band
at one of the TRIM $\bm{\Gamma}_{n_1n_2n_3}=(n_1\bm{G}_1+n_2\bm{G}_2+n_3\bm{G}_3)/2$.
$\bm{G}_{i=1,2,3}$ are primitive reciprocal lattice vectors.
When $\nu _0=1$,
the system is guaranteed to be in the NLS phase.
When the SOC is taken into account,
the $\mathbb{Z}_2$ invariants coincide with Fu--Kane--Mele invariants \cite{Fu07, Kim15}.
In fact, the SOC changes an NLS into an insulator or a Dirac semimetal,
which has corresponding Fu--Kane--Mele invariants. \cite{Kim15, Kobayashi}

To intuitively understand the appearance of a nodal line in 
a spinless system with TR and inversion symmetries, it 
is useful to consider the following effective model. 
An effective model for a single valence band and a single conduction band with these symmetries 
is 
generally expressed as 
\begin{align}
&H({\bf k})=a_0({\bf k})+a_x({\bf k})\sigma_x+a_z({\bf k})\sigma_z,
\end{align}
where a $\sigma_y$ term is prohibited by these symmetries. 
Its 
band gap closes
only if the two conditions $a_x({\bf k})=0$ and $a_z({\bf k})=0$ are satisfied simultaneously.
Their solutions give a nodal line 
in ${\bf k}$ space, and the Berry phase $\phi (\ell)$ 
around the nodal line
turns out to be $\pi$, 
in agreement with the discussion in the previous paragraph.

In many NLSs, characteristic surface states called 
drumhead surface states appear. In the surface Brillouin zone, they appear in 
a region surrounded by the projection of the nodal line (Fig.~\ref{fig:nodalline}(a)).
While the appearance of drumhead surface states can be seen from a calculation 
based on an effective model \cite{Kim15}, they do not
necessarily appear in every NLS \cite{Hirayama-Ca}.

\begin{figure}[ptb]
\begin{center}
\includegraphics[clip,width=0.45\textwidth ]{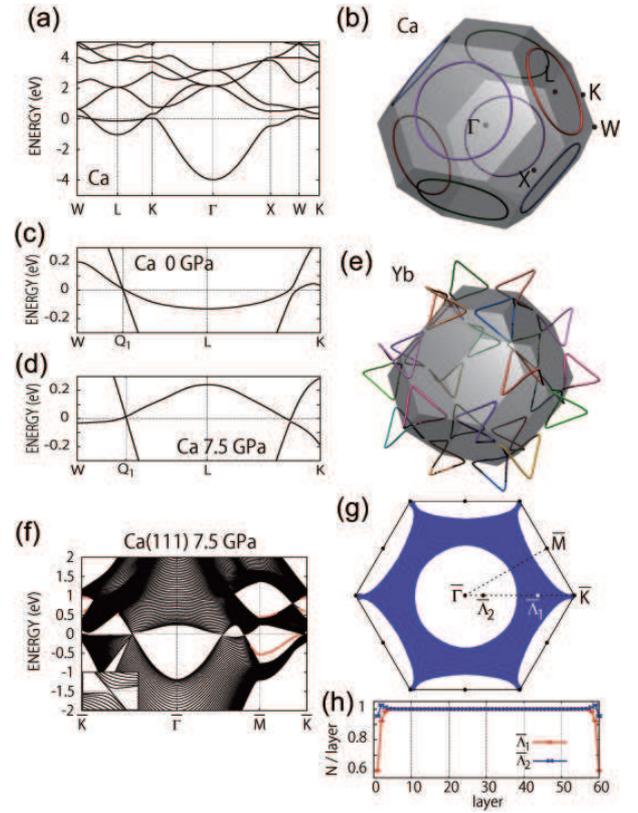} 
\caption{(Color online) 
Electronic band structure and nodal lines in alkaline earth metals.
(a) Electronic band structure of Ca within the LDA. 
(b) Nodal lines and the Brillouin zone of Ca, where identical nodal lines are shown in the same color. 
(c) (d) Magnified electronic band structure of Ca at ambient pressure and 
at 7.5 GPa, respectively, within the LDA. 
(e) Nodal lines and the Brillouin zone of Yb when the SOC is neglected.
(f) Electronic band structure for the (111) surface of Ca within the LDA, with 
states close to the nodal line magnified in the inset. The bands in red represent 
states which monotonically decay into the bulk region.
(g) Zak phase at the surface momenta ${\bf k}_{\parallel}$ in the (111) surface Brillouin zone. The shaded region represents ${\bf k}_{\parallel}$  with the $\pi$ Zak phase, while other regions represent ${\bf k}_{\parallel}$  with the $0$ Zak phase. 
(h) Charge profile for two values of the surface momentum ${\bf k}_{\parallel}=\bar{\Lambda}_1,\bar{\Lambda}_2$
in real space in the thickness direction of the slab. 
The vertical axis represents the charge density per surface unit cell in each atomic layer, measured in units of  electronic charge, $-e$.
The energy is measured from the Fermi level.
(From Nat. Commun. 8, 14022 (2017) \cite{Hirayama-Ca}. Reprinted with permission from Springer Nature.)}
\label{fig:Ca}
\end{center}
\end{figure}

Nodal lines from this mechanism can be seen in 
alkaline-earth metals Ca, Sr, and in Yb when the SOC
is neglected, from {\it ab initio} calculation.
In reality, the SOC gives 
a small gap along the otherwise gapless nodal lines. 
Ca, Sr, and Yb are nonmagnetic metals, having a face-centered cubic (fcc) lattice, with the 
space group Fm$\bar{3}$m (No.225). Let $a$ denote the lattice parameter. 
Figure~\ref{fig:Ca}(a) shows the electronic structure of Ca
obtained by local density approximation (LDA).
The top of the valence band, which is relatively flat near the L points, originates from the $4p$ orbital oriented along the [111] axis having strong $\sigma $ bonding, while most of the 
other valence bands originate from the $4s$ and $3d$ orbitals.
Around the L points, their crossing yields four
nodal lines around the L points within approximately $\pm$ 0.01 eV near the Fermi level, 
as shown in Fig.~\ref{fig:Ca}(a). 
The four nodal lines (Fig.~\ref{fig:Ca}(b)) are slightly away from the faces of the first Brillouin zone, 
except along the $L$-$W$
lines (Q$_1$ in Fig.~\ref{fig:Ca}(c)) by $C_2$ symmetry.
The Berry phase around each nodal line is numerically confirmed to be $\pi$.
At ambient pressure, Ca is not an NLS, because 
the two bands forming the nodal lines both disperse downward 
around the L points (Fig.~\ref{fig:Ca}(c)). 
Ca becomes an NLS under pressure, 
as shown in the band structure at 7.5 GPa  in Fig.~\ref{fig:Ca}(d).
The nodal lines in Yb with the neglected SOC are shown in 
Fig.~\ref{fig:Ca}(e). These nodal lines can be understood from those in Ca via a Lifshitz transition of 
nodal lines. 
Figure~\ref{fig:Ca}(f) shows the electronic structure of the Ca(111) surface at 7.5 GPa.
Drumhead surface states connecting the gapless points exist near the Fermi level around 
the $\bar{\rm M}$ point, which are isolated from the bulk states.

\subsection{Surface charges and nodal lines with the $\pi$ Berry phase}

This $\pi$ Berry phase is closely related to the Zak phase, as we explain here. 
We calculate the Zak phase
along a certain reciprocal vector ${\bf G}$.
For the calculation, we decompose the wavevector ${\bf k}$ into the components
along ${\bf n}\equiv {\bf G}/|{\bf G}|$ and perpendicular to ${\bf n}$:
${\bf k}=k_{\perp}{\bf n}+{\bf k}_{\|}$, ${\bf k_{\|}}\perp{\bf n}$. 
For each value of  ${\bf k}_{\parallel} $,  one can define
the Zak phase by
\begin{align}
\theta ({\bf k}_\parallel )=-i\sum_{n}^{{\rm occ.}}\int _{0 }^{|{\bf G}|} d k_{\perp} 
\left\langle{u_n({\bf k})}\right| \nabla_{k_{\perp}} \left|{u_n({\bf k})}\right\rangle,
\label{eq:Zak}\end{align}
where 
 $u_n({\bf k})$ is the bulk eigenstate in the $n$th band and the sum is over the occupied states. 
Here, the gauge is taken as
$u_n({\bf k})=u_n({\bf k}+{\bf G})e^{i{\bf G}\cdot{\bf r}}$.
The Zak phase is defined in terms of modulo $2\pi$. 
We focus on the cases without the SOC and 
neglect the spin degeneracy. Under both inversion and TR symmetries,
the Zak phase takes a quantized value of 0 or $\pi \pmod{2\pi}$ \cite{Chan15,PhysRevB.88.245126}.
The Zak phase is related to the charge polarization at surface momentum ${\bf k}_{\|}$ for a surface perpendicular 
to ${\bf n}$ \cite{PhysRevB.48.4442,resta}.
Additionally, in a 3D system, if the system at each ${\bf k}_{\|}$ is regarded 
as a 1D system, then the product of 
the Zak phase $\theta({\bf k}_{\|})$ and $e/(2\pi)$
is equal to $\sigma({\bf k}_{\|})$ modulo $e $, where $\sigma({\bf k}_{\|})$ is the surface charge when the system at 
a given ${\bf k}_{\|}$ is regarded as a 1D system. For an 
insulator, the surface polarization charge density $\sigma_{\rm total}$  at a given surface is
given by $\sigma_{\rm total}=\int\frac{d^2k_{\|}}{(2\pi)^2}\sigma({\bf k}_{\|})$ \cite{PhysRevB.48.4442}.

Because the Berry phase around the nodal line is $\pi$,  
the Zak phase jumps by 
$\pi$ as ${\bf k}_{\|}$ changes across the nodal line.
For example, for the Ca (111) surface, the region of the $\pi$ Zak phase is shown as the shaded region
in Fig.~\ref{fig:Ca}(g), and this region is surrounded by the projections of the nodal lines.  
For example, because 
the Zak phases for points $\bar{\Lambda}_1$ and $\bar{\Lambda}_2$ in Fig.~\ref{fig:Ca}(h) 
are  $\pi$ and $0$, respectively,  
the surface polarization charges are $\sigma(\bar{\Lambda}_1)\equiv e/2$ and $\sigma(\bar{\Lambda}_2)\equiv 0$  (mod $e$).
This difference in surface polarization charges 
can be directly seen in the charge distribution for the slab geometry.
At $\bar{\Lambda}_2$ the charge distribution is almost constant, whereas at $\bar{\Lambda}_1$ it decreases 
by $\sim (-e)/2$ near each surface.
Thus, the $e/2$ surface polarization charge from the $\pi$ Zak phase
is attributed to
bulk states, not to surface states \cite{Hirayama15}. 

The area of the region with $\sigma({\bf k}_{\|})=\frac{e}{2}$ in Fig.~\ref{fig:Ca}(g) 
is $0.485$ of the area of the Brillouin zone, and the total  
surface polarization charge density is $0.485\cdot \frac{e}{2}
\sim 0.243e$ per surface unit cell. 
A nonzero 
surface polarization charge does not violate the inversion symmetry because 
the two surfaces of the slab have the same surface polarization charges.
Nevertheless, in reality this amount of surface charge does not appear since the correspondence between 
the surface charge and the Zak phase is correct only for insulators. 
In reality, the excess surface charge is screened by 
free carriers because the system is a semimetal, and 
this results in surface dipoles at
the surface. 
From the above argument, 
carriers in the semimetal screen the surface charge within a screening length on the order of nanometers. 

\subsection{Transitions from nodal-line semimetals to other topological phases}
\begin{figure}[ptb]
\begin{center}
\includegraphics[clip,width=8cm]{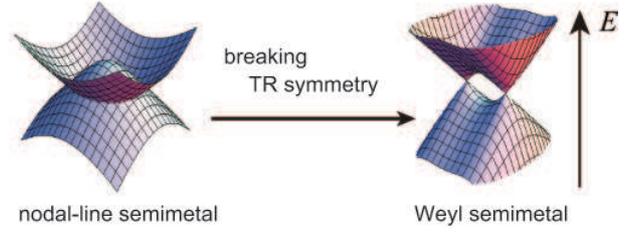} 
\caption{(Color online) \label{NLStoWeyl}
Schematic picture of the transition from an NLS to a
Weyl semimetal by breaking the TR symmetry.
}
\end{center}
\end{figure}

Thus far, we discussed NLSs with the $\pi$ Berry phase. Because 
this NLS phase requires both TR and inversion symmetries, 
breaking of either TR or inversion symmetry leads to transitions into other phases 
including topological semimetal phases. 
As an example, let us consider a nodal line encircling one of the TRIM, such as one of those in Ca (Fig.~\ref{fig:Ca}(b)). By breaking  TR symmetry, such an NLS
always becomes a spinless WSM \cite{Okugawa17}, which 
can be easily seen in terms of an effective model. 
The nodal line is formed by the conduction and valence bands with opposite parity eigenvalues at the TRIM. \cite{Okugawa17}
Therefore, the parity operator is given by $P=\sigma _z$, and
the nodal line can be described by the effective Hamiltonian
\begin{align}
H(\bm{q})=a_0(\bm{q})+a_y(\bm{q})\sigma _y+a_z(\bm{q})\sigma _z, \label{effH}
\end{align}
where $\bm{q}$ is the wavevector measured from the TRIM.
The nodal line is described as $a_y(\bm{q})=a_z(\bm{q})=0$.
A TR-breaking term is then represented by $a_x(\bm{q})\sigma _x$,
with $a_x(\bm{q})$ being an odd function of $\bm{q}$ 
owing to inversion symmetry.
Since the nodal line encloses the TRIM ($\bm{q}=0$) by assumption,
wavevectors satisfying $a_x(\bm{q})=0$ always exist somewhere on the nodal line, and 
Weyl nodes necessarily emerge on it.
As a result, when a nodal line encloses one of the TRIM,
the NLS
undergoes a transition to
a spinless WSM by breaking TR symmetry \cite{Okugawa17}, as schematically shown in 
Fig.~\ref{NLStoWeyl}.
It is theoretically predicted that this kind of phase transition 
is driven by circularly polarized light in electronic systems. \cite{Narayan16, Yan16, Chan16, Taguchi16, Ezawa17, Yan17}.
Meanwhile, when nodal lines do not enclose one of the TRIM, the TR breaking does not necessarily lead to topological semimetals, but it may lead either to an insulator or to a Weyl semimetal.
Similarly, 
by breaking the inversion symmetry, the NLS either remains an NLS or  
becomes an insulator or a Weyl semimetal, 
depending on crystallographic symmetries \cite{Okugawa17}.
Thus, by starting with one topological semimetal material, one can obtain other topological phases
by breaking symmetries, and such topological phase transitions stem from the interplay between 
topology and symmetry. Understanding of this interplay will give further insight in  
{\it ab initio} studies of topological semimetal materials. 

\section{Numerical Tools for Characterizing Topological Semimetal Materials}
In this section, for readers' convenience we briefly summarize the numerical methods used in this review. 

\subsection{Self-energy correction}
In general, the band gap estimated in the density functional theory (DFT) with the LDA  
is smaller than its real value. 
Therefore, the LDA tends to predict nontrivial phases in narrow- and pseudogap systems
even when they are in the trivial phases.
Thus, high precision in {\it ab initio} calculations is required to search for nontrivial topological phases of matter. 
One of the suitable approaches beyond the LDA is the Green's function method for self-energy correction. 
Within the Green's function method, the GW method is the most widely used approximation for the self-energy~\cite{hedin65}. 
The band structure in the GW approximation (GWA) is obtained by diagonalizing the following GW Hamiltonian: 
\begin{equation}
\mathcal{H}^{\text{$GW$}} = \mathcal{H}^{\text{LDA}} -V_{\text{xc}}^{\text{LDA}}+\Sigma ^{GW},
\label{eq:gw}
\end{equation}
where $\mathcal{H}^{\text{LDA}}$, $V_{\text{xc}}^{\text{LDA}}$, and $\Sigma ^{GW}$ are 
the Kohn-Sham Hamiltonian in the LDA, the LDA exchange-correlation potential, and the self-energy in the GWA, respectively.
$\Sigma ^{GW}$ is obtained from the convolution integral between the Green's function $G$ in the LDA and the screened interaction $W$. 
For example, within the LDA, tellurium (Te) is incorrectly described as a metal in the LDA, and 
the GW method correctly reproduces the existence of the band gap.

\subsection{Wannier function}
In discussing the topology of electronic structures,
it is useful to derive an \textit{ab initio} low-energy effective model describing the bands near the Fermi level.
The transfer integral in the low-energy effective model is given by 
\begin{equation}
t_{mn}(\bm{R})= \langle \varphi _{m\bm{0}}|\mathcal{H}| \varphi_{n\bm{R}} \rangle,
\label{eq:t}
\end{equation}
where $\varphi_{m\bm{R}}$ is the maximally localized Wannier function (MLWF) of the $m$th orbital localized at the unit cell $\bm{R}$ in the space 
spanned by the states near the Fermi energy~\cite{marzari97,souza01}.
The MLWF  is defined from the associated Bloch function
\begin{equation}
|\varphi _{m}  (\bm{k}) \rangle =\sum _n U_{mn}(\bm{k})| u _{n} (\bm{k}) \rangle ,
\end{equation}
where $\varphi _{m}  (\bm{k})$ is the Fourier transform of $\varphi _{m\bm{R}}$
and the coefficients $U_{mn}(\bm{k})$ are determined such that the quadratic extent of the wave function
\begin{equation}
V  =\sum _m(\langle \varphi _{m\bm{0}}|\bm{r}^2|\varphi _{m\bm{0}}\rangle -|\langle \varphi _{m\bm{0}}|\bm{r}|\varphi _{m\bm{0}}\rangle |^2)
\label{eq:omega}
\end{equation}
is minimized.
The position operator in Eq. (\ref{eq:omega}) is treated in terms of  the Berry connection in ${\bf k}$ space.

\subsection{Monopole charge}
In the search for Weyl semimetals using {\it ab initio} calculations, 
it is sometimes difficult to see whether there is a Weyl node or not because it is 
numerically difficult to distinguish between the case with no gap and that with a tiny gap. 
This problem is solved by numerically calculating the value of the monopole charge at the ${\bf k}$ point considered.
By extending the numerical scheme for the calculation of the Chern number \cite{fukui05},
we can calculate the monopole charge by the following scheme. 

\begin{figure}[ptb]
\begin{center}
\includegraphics[clip,width=8cm]{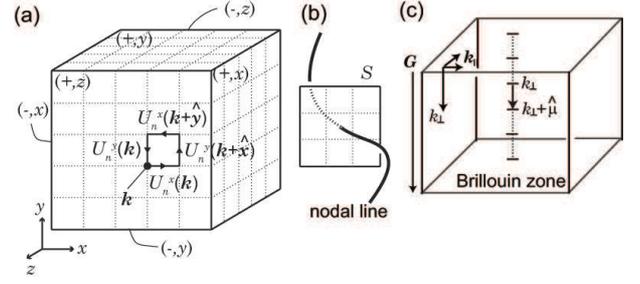} 
\caption{\label{BerryCal}
(a) Schematic picture of a numerical calculation of the monopole charge in a cubic region.
$\hat{\bm{x}}$ and  $\hat{\bm{y}}$ are the vectors between the nearest $\bm{k}$-mesh points on the $(+,z)$ surface of the
cube.
(b) Schematic picture of a numerical calculation of the Berry phase around the nodal line.
(c) Schematic picture of a numerical calculation of the Zak phase along the reciprocal lattice
vector $\bm{G}_{\perp}$.
}
\end{center}

\end{figure}

We first define a small cubic region around the ${\bf k}$ point, as shown in Fig.~\ref{BerryCal}(a), and calculate 
the integral of the Berry curvature over the surfaces of this region. 
To this end, the faces of this cubic region are divided into small square meshes. 
For each edge of the small square meshes, we define a $U(1)$ link variable for the $n$th band as
\begin{equation}
U_{n}^{\mu}(\bm{k}) = \frac{\langle u_n (\bm{k})|u_n (\bm{k}+\hat{\bm{\mu}}) \rangle}{
|\langle u_n (\bm{k})|u_n (\bm{k}+\hat{\bm{\mu}}) \rangle |},
\label{eq:umu2}
\end{equation}
where
$\mu\ (=x,y,z)$ is the direction of the edge and
$\hat{\bm{\mu}}$ is the vector along the edge between the nearest mesh points along the $\mu$ direction.
We label the six faces of the cubic region as $(s,\nu)$ according to the orientations of their normal vectors $s\nu=\pm\nu$ out of the cube, where $\nu=x,y,z$
and $s=\pm$. 
The monopole charge $\tilde{c}_n$ in the cube is calculated from $U_n^{\mu}(\bm{k})$ as
\begin{align}
\tilde{c}_{n} &=\sum_{s=\pm}\sum_{\nu=x,y,z}\tilde{c}_{n}^{(s,\nu)},\\
\tilde{c}_{n}^{(s,\nu)}&= \frac{1}{2\pi i}s\sum_{\bm{k}\in S_{s,\nu}}\varepsilon_{\nu\mu\rho}
\left\{\tilde{F}_n^{\mu\rho}(\bm{k})
-\Delta^\mu \tilde{A}_n^\rho(\bm{k})+\Delta^\rho \tilde{A}_n^\mu(\bm{k})\right\},
\label{eq:cnn}
\end{align}
where $\bm{k}$ is a mesh point on the face specified by $(s,\nu)$, so that the square
meshes with vertices ($\bm{k}$, $\bm{k}+\bm{\mu}$, $\bm{k}+\bm{\mu}+\bm{\rho}$, $\bm{k}+\bm{\rho}$) cover the face.
$\tilde{F}_n^{\mu\rho}(\bm{k})$ is the lattice field strength in the square mesh 
 ($\bm{k}$, $\bm{k}+\bm{\mu}$, $\bm{k}+\bm{\mu}+\bm{\rho}$, $\bm{k}+\bm{\rho}$) on the faces of the cube:
\begin{align}
\tilde{F}_n^{\mu\rho}(\bm{k}) &= \ln U_n^\mu(\bm{k})U_n^\rho(\bm{k}+\hat{\bm{\mu}})U_n^\mu(\bm{k}+\hat{\bm{\rho}})^{-1}U_n^\rho(\bm{k})^{-1}  ,
\label{f12}\\
-\pi &< \frac{1}{i}\tilde{F}_n^{\mu\rho}(\bm{k}) \leq  \pi.
\label{eq:f12rng}
\end{align}
$\Delta ^{\mu} \tilde{A}_n^{\rho}(\bm{k})$ is the difference in the gauge potential between the neighboring 
mesh points displaced by $\hat{\bm{\mu}}$:
\begin{align}
\Delta ^{\mu} \tilde{A}_n^{\rho}(\bm{k}) &=\tilde{A}_n^{\mu}(\bm{k}+\hat{\bm{\mu}})
-\tilde{A}_n^{\mu}(\bm{k}),
\label{eq:amu}\\
\tilde{A}_n^{\mu}(\bm{k})&=\ln U_n^{\rho}(\bm{k}) ,\ -\pi < \frac{1}{i} \tilde{A}_n^{\mu}(\bm{k}) \leq  \pi .
\label{eq:amurng}
\end{align}
The numerical convergence of this method with increasing the number of meshes is very rapid.
Several tens of mesh points per each direction is usually sufficient to
obtain a correct result.

\subsection{Berry phase}
The nodal line discussed in Sect.~\ref{sec:nodalline} is characterized by the $\pi$ Berry phase around the
nodal line. Here we explain the numerical method used to calculate this Berry phase around the nodal line. 
We consider a square encircling the nodal line, and we divide the square into meshes  (Fig.~\ref{BerryCal}(b)). 
For notational simplicity, we set the edges of the square to be along the $x$ and $y$ axes.
The Berry phase for the $n$th band along the square is defined as
\begin{equation}
\phi  _n = \sum_{\bm{k} \in S}[\tilde{F}_n^{xy}(\bm{k})
-\Delta^x \tilde{A}_n^y(\bm{k})+\Delta^y \tilde{A}_n^x(\bm{k})] ,
\label{eq:bpn}
\end{equation}
where $\tilde{F}_n^{xy}(\bm{k})$ and $\Delta ^{\mu} \tilde{A}_n^{\nu}(\bm{k})$ are the lattice field strength and 
the difference in the gauge potential for the meshes in the square $S$, and the summation is taken so that the
meshes
cover the whole square.
In spinless systems with TR and inversion symmetries, the Berry phase is 
quantized to be either $0$ or $\pi$ (mod $2\pi$), and 
if it is $\pi$, it is indeed the nodal line discussed in Sect.~\ref{sec:nodalline}.

\subsection{Zak phase}
The Zak phase for $\bm{k}_{\parallel }$ is calculated as~\cite{smith93,PhysRevB.48.4442,resta}
\begin{equation}
\theta (\bm{k}_{\parallel })=\text{Im}  \biggl\{ \text{In}  \prod_{\bm{k}_{\perp }}  \text{det} (\langle  u_m(\bm{k}_{\parallel },\bm{k}_{\perp }) | u_n(\bm{k}_{\parallel },\bm{k}_{\perp }+\hat{\bm{\mu }} ) \rangle )  \biggl\},
\label{eq:zpn}
\end{equation}
where $m$ and $n$ in the determinant run over the occupied bands.
$\hat{\bm{\mu}}$ is the vector between the nearest mesh points along the reciprocal lattice vector $\bm{G}$,
and the product is taken over the wavevector $\bm{k}$ such that the path covers  
the reciprocal lattice vector $\bm{G}$, as shown in Fig.~\ref{BerryCal}(c).
Note that in calculating the Zak phase, the gauge should be chosen to satisfy
$|u_m(\bm{k}_{\perp })\rangle=e^{i\bm{G}\cdot\bm{r}}|u_m(\bm{k}_{\perp }+\bm{G})\rangle$.
This gauge choice is related to the choice of the unit cell, and it affects the value of the Zak phase and the polarization \cite{PhysRevB.48.4442}.

\section{Summary}
We have seen various types of topological semimetals, which have 
degeneracies between bands, 
apart from those originating from symmetry.
After brief reviews of basic properties of topological semimetals, we
explained how these properties are used for the search for topological semimetals
using {\it ab initio} calculations.
For example, we showed that LuSI at high pressure and Te at high pressure are Weyl semimetals, 
and that Ca under pressure and HfS are nodal-line semimetals.
We also explained that various topological metals are related to each other in 
a complex way.
This shows a profound interplay between topology and symmetry in the band theory of solids.
Thus, the studies on TIs and topological metals 
have provided us with renewed interest in the band theory of solids.

Through close collaborations among theories, experiments, and {\it ab initio}
calculations, we have been obtaining new insights into the electronic properties of solids.




\begin{acknowledgments}
We thank Takashi Miyake for fruitful discussions and Shoji Ishibashi for providing us with the {\it ab initio} code (QMAS) and pseudopotentials.
This work was supported by Grants-in-Aid for Scientific Research
(Nos.26287062, 26600012);
by the Computational Materials Science Initiative (CMSI), 
Japan; by CREST, JST (Grant No. JPMJCR14F1);  
by JSPS KAKENHI Grant Number 16J08552;
and also by MEXT Elements Strategy Initiative to Form Core Research Center (TIES).
\end{acknowledgments}


\end{document}